\documentclass[twocolumn,showpacs,preprintnumbers,prl,aps,superscriptaddress]{revtex4-2}
\usepackage{graphicx}
\pdfoutput=1
\usepackage{amsmath}
\usepackage{amssymb}
\usepackage[retain-unity-mantissa=false]{siunitx}
\usepackage{xcolor}
\usepackage[caption=false]{subfig}
\usepackage{placeins}
\usepackage[colorlinks=true, linkcolor=blue, citecolor=blue, filecolor=blue, urlcolor=blue,pdfproducer={.},pdfcreator={.}]{hyperref}
\usepackage{cleveref}
\usepackage{blindtext}
\usepackage{comment}
\usepackage{verbatim}
\usepackage{ulem}
\usepackage[utf8]{inputenc}
\usepackage[T1]{fontenc} 
\usepackage{lmodern} 
\usepackage{physics}
\usepackage{booktabs} 
\usepackage{ragged2e} 
\usepackage{array} 
\usepackage{float}
\usepackage{tikz}
\usetikzlibrary{decorations.pathmorphing}
\usetikzlibrary{decorations.text}
\usetikzlibrary{calc}
\usepackage{svg}
\usepackage{natbib}

\begin{document}

\title{Spin transport and magnetic proximity effect in CoFeB/normal metal/Pt trilayers}

\author{\underline{Simon~\surname{Häuser}}}
    \email{shaeuser@rptu.de}
    \affiliation{Department of Physics and Research Center OPTIMAS, Rheinland-Pfälzische Technische Universität Kaiserslautern-Landau, 67663 Kaiserslautern, Germany}
    
    \author{\underline{Matthias~R.~\surname{Schweizer}}}
    \email{mschweiz@rptu.de}
    \affiliation{Department of Physics and Research Center OPTIMAS, Rheinland-Pfälzische Technische Universität Kaiserslautern-Landau, 67663 Kaiserslautern, Germany}
    
    \author{Sascha \surname{Keller}}
    \affiliation{Department of Physics and Research Center OPTIMAS, Rheinland-Pfälzische Technische Universität Kaiserslautern-Landau, 67663 Kaiserslautern, Germany}
    
    \author{Andres \surname{Conca}}
    \affiliation{Department of Physics and Research Center OPTIMAS, Rheinland-Pfälzische Technische Universität Kaiserslautern-Landau, 67663 Kaiserslautern, Germany}
    \affiliation{Instituto de Micro y Nanotecnología, IMN-CNM, CSIC (CEI UAM+CSIC), 28760 Tres Cantos, Madrid, Spain}
    
    \author{Moritz \surname{Hofherr}}
    \affiliation{Department of Physics and Research Center OPTIMAS, Rheinland-Pfälzische Technische Universität Kaiserslautern-Landau, 67663 Kaiserslautern, Germany}
    
    \author{Evangelos \surname{Papaioannou}}
    \affiliation{Department of Physics and Research Center OPTIMAS, Rheinland-Pfälzische Technische Universität Kaiserslautern-Landau, 67663 Kaiserslautern, Germany}
    \affiliation{Department of Physics, Aristotle University of Thessaloniki, 54124 Thessaloniki, Greece}

    \author{Benjamin \surname{Stadtmüller}}
    \affiliation{Department of Physics and Research Center OPTIMAS, Rheinland-Pfälzische Technische Universität Kaiserslautern-Landau, 67663 Kaiserslautern, Germany}
    \affiliation{Institute of Physics, Johannes Gutenberg University Mainz, 55128 Mainz, Germany}

    \author{Burkard \surname{Hillebrands}}
    \affiliation{Department of Physics and Research Center OPTIMAS, Rheinland-Pfälzische Technische Universität Kaiserslautern-Landau, 67663 Kaiserslautern, Germany}
    
    \author{Martin \surname{Aeschlimann}}
    \affiliation{Department of Physics and Research Center OPTIMAS, Rheinland-Pfälzische Technische Universität Kaiserslautern-Landau, 67663 Kaiserslautern, Germany}
    
    \author{Mathias \surname{Weiler}}
    \affiliation{Department of Physics and Research Center OPTIMAS, Rheinland-Pfälzische Technische Universität Kaiserslautern-Landau, 67663 Kaiserslautern, Germany}

\date{\today}
\begin{abstract}

We present a study of the damping and spin pumping properties of CoFeB/X/Pt systems with $\rm X=Al,Cr$ and $\rm Ta$. We show that the total damping of the CoFeB/Pt systems is strongly reduced when an interlayer is introduced independently of the material. Using a model that considers spin relaxation, we identify the origin of this contribution in the magnetically polarized Pt formed by the magnetic proximity effect (MPE), which is suppressed by the introduction of the interlayer. The induced ferromagnetic order in the Pt layer is confirmed by transverse magneto-optical Kerr spectroscopy at the M$_{2,3}$ and N$_7$ absorption edges as an element-sensitive probe. We discuss the impact of the MPE on parameter extraction in the spin transport model. 
 
\end{abstract}

\maketitle

\section{Introduction}

Spin pumping~\cite{tser, tser2}, describes the increased magnetic damping when the magnetization of a ferromagnetic (FM) layer in contact with a non-magnetic (NM) layer is excited by a microwave field under ferromagnetic resonance conditions. A spin current is generated by the precessing magnetization and injected into the NM layer where it undergoes spin-flip processes leading to the increase in damping. The transparency of the FM/NM interface, which determines the magnitude of the injected spin current, is governed by the spin mixing conductance $G^{\uparrow\downarrow}$~\cite{tser}. When heavy metals, in particular Pt, are used as NM, a large spin pumping effect is observed due to the large spin-orbit interaction that suppresses spin backflow. The magnetic proximity effect (MPE), which induces a static magnetic polarization in Pt films in direct contact with a metallic FM~\cite{caminale,suzuki,wilhelm,kuschel2016,antel1999}, can have a profound impact on the spin pumping damping~\cite{caminale}. However, the impact of the MPE on material parameters extracted from spin transport models used to describe the spin pumping process is not very well understood.

Here, we experimentally study the impact of the MPE on spin pumping in FM/NM/Pt trilayers and evaluate how the MPE affects parameters extracted from a phenomenological spin transport model~\cite{Boone2015}. 

CoFeB/Pt is a good model system to study this topic due to the low damping properties of CoFeB ~\cite{bilzer,liu,cofeb-ann, cofeb-alpha}. Recently, we reported a significant lack of correlation between the measured voltages and the effective value of $G^{\uparrow\downarrow}$ when comparing CoFeB/Pt and CoFeB/Ta, which can only be explained by interface effects~\cite{lack}. \\
Therefore, we investigate CoFeB/X/Pt systems with $\rm X=Al,Cr,Ta$ to modify the interface and suppress the MPE. We experimentally verify the presence (absence) of MPE in CoFeB/Pt bilayers and reference samples using a table-top high harmonic extreme UV radiation source. We reproduce the damping parameters of the CoFeB/X/Pt systems by using an effective spin transport model~\cite{Boone2015}. We find that the effective spin diffusion length of Pt is reduced for sample systems with MPE.

\section{Experimental details}

We use polycrystalline Cr(7.5)/MgO(4.7)/CoFeB(11)/
X(t)/Pt,Ta(3) systems, where $\rm CoFeB=Co_{40}Fe_{40}B_{20}$ and $\rm X=Al,Cr,Ta$, grown by RF sputtering on Si substrates passivated with SiO$_2$. Thickness is given in nm. No annealing was performed. A microstrip-based VNA-FMR setup was used to study the damping properties. A detailed description of the FMR measurement and analysis procedure is given in previous work~\cite{fept,cofeb-ann}. 

The magnetic order in the multilayer films was investigated by transverse magneto-optical Kerr spectroscopy (HHG-TMOKE) using fs-XUV radiation obtained by high harmonic generation~\cite{rundquist1998}. A detailed description of the HHG-TMOKE setup can be found in~\cite{mathias2012}. The element-resolved magnetic contrast of pure CoFeB/Pt, CoFeB/MgO and CoFeB/Al/Pt is compared to prove the existence of induced ferromagnetic order in Pt due to MPE.

\section{Direct confirmation of MPE}

\begin{figure}[h]
    \includegraphics[width=1\columnwidth]{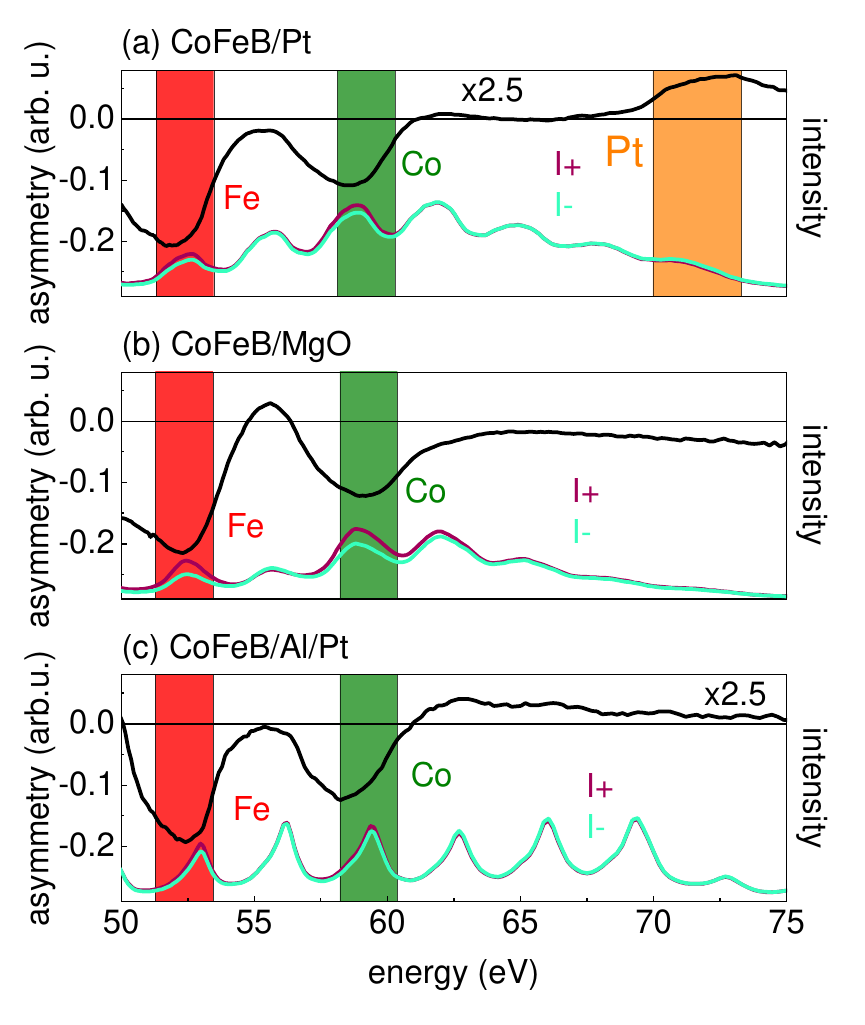}
	  \caption{\label{mpe}(Black) Static magnetic asymmetry of the CoFeB/Pt bilayer film (a) and a reference CoFeB film capped with MgO in (b). In (c) the static magnetic asymmetry of a CoFeB/Al(4)/Pt multilayer is shown. The colored shaded areas represent the M$_{2,3}$ and N$_7$ absorption edges of the elements. In addition, the measured reflected XUV spectra I$_+$ and I$_-$ are shown. Note that the asymmetries of CoFeB/Pt and CoFeB/Al(4)/Pt have been multiplied by a factor of 2.5 for better comparability with the results for pure CoFeB. The clear signal of an induced Pt magnetic moment due to the MPE is observed only in (a).}
\end{figure}

Using high harmonic generation (HHG) extreme ultraviolet pulses as an element-sensitive probe~\cite{vorakiat2009}, we investigate the magnetic properties of CoFeB/Pt and reference CoFeB/MgO and CoFeB/Al/Pt samples. The magnetic contrast is obtained by calculating the magnetic asymmetry $(I_+-I_-)/(I_++I_-)$. Here, I$_+$ and I$_-$ denote the spectrally resolved reflected intensities of the XUV probe pulse from the sample surface for two opposing external and alternately applied magnetic fields. The results are shown in Fig.~\ref{mpe}. Note that the amplitudes of Fe and Co within the static asymmetries of CoFeB/Pt and CoFeB/Al/Pt are lower by a factor of about 2.5 compared to the pure CoFeB sample due to the Pt and Al/Pt layers covering the CoFeB.\\
All samples show clearly enhanced magneto-optical contrast near the M$_{2,3}$ absorption edges of Fe ($\sim$53 eV) and Co ($\sim$59 eV)~\cite{xraybooklet}. However, only for CoFeB/Pt an additional magnetic asymmetry can be observed for photon energies at the Pt N$_7$ absorption edge at $\sim$71.2 eV. This observation is consistent with the formation of a magnetically ordered phase in Pt due to the MPE. 

\section{Damping Measurements}

\begin{figure}[h]
    \includegraphics[width=1\columnwidth]{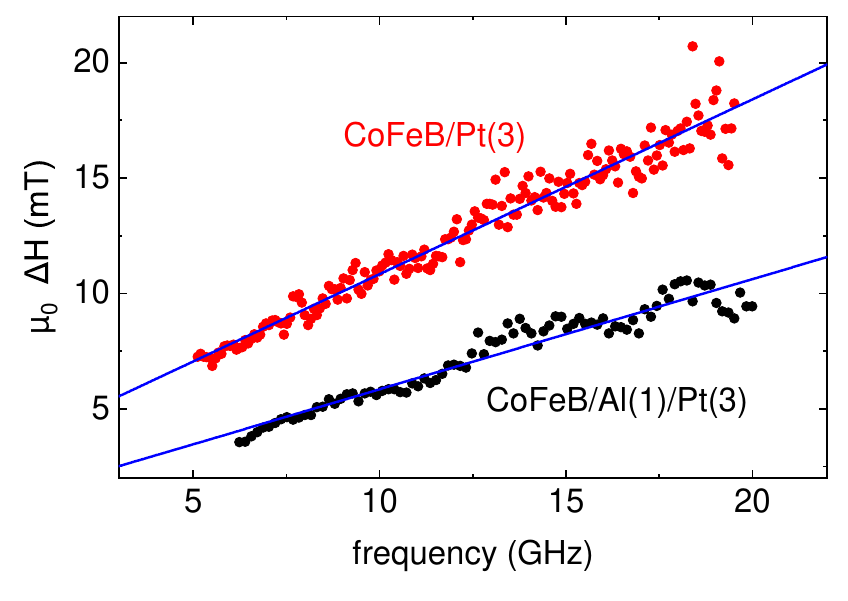}
	  \caption{\label{linewidth}Frequency dependence of the FMR linewidth $\Delta H$ for CoFeB/Pt and a CoFeB/Al/Pt system. The blue lines are fits to Eq.~(\ref{gilbert}) to extract the damping parameter $\alpha$.}
\end{figure}

Fig.~\ref{linewidth} shows the frequency dependence of the FMR linewidth for a CoFeB/Pt and a CoFeB/Al(1)/Pt system. We fit the data to 
\begin{equation} \label{gilbert}
\mu_0\Delta H= \mu_0\Delta H_0 + \frac{4 \pi \alpha f}{\gamma}\,
\end{equation}
to extract the damping parameter $\alpha$ from the slope of the linear fits (blue lines).
Here, $\Delta H_0$ is the inhomogeneous broadening and is related to the film quality. The CoFeB/Pt bilayer shows a relatively large damping of $\alpha=(10.5 \pm 0.8)\times 10^{-3}$, which is in good agreement with previously reported data~\cite{lack}. The obtained value is the result of several contributions. In addition to the characteristic damping of CoFeB ($\alpha_0$), a loss of angular momentum occurs due to the generation of a spin current and its injection into Pt via the spin pumping process, resulting in a subsequent damping increase ($\alpha_{\rm sp}$). 
In addition, the presence of ferromagnetic order in Pt near the interface due to the magnetic proximity effect also increases the effective damping ($\alpha_{\rm MPE}$). Recently we have shown that this contribution is strong in CoFeB/Pt~\cite{lack}.

Fig.~\ref{linewidth} shows the damping reduction by introducing a \SI{1}{\nano\meter} thick Al interlayer. The effective damping parameter is strongly reduced to $\alpha=(6.6 \pm 0.2)\times 10^{-3}$. The separation of the Pt layer from the direct contact with the ferromagnetic layer suppresses the MPE and the impact of the MPE on the damping vanishes. However, in a first approximation considering the interfaces only as passive elements, the contribution ($\alpha_{\rm sp}$) is expected to remain unchanged because the spin diffusion length of Al exceeds 10\,nm~\cite{Bass2007}.

\begin{figure}[h]
  \centering
  \includegraphics[width=1\columnwidth]{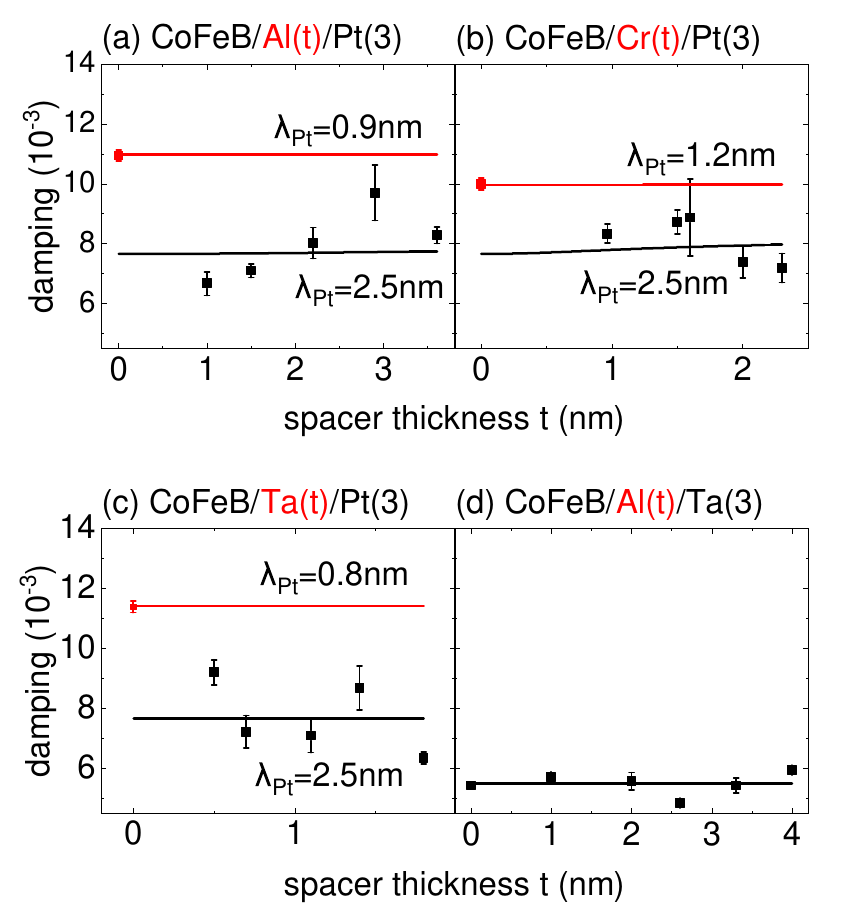}
    \caption{Dependence of the measured damping parameter $\alpha$ on the interlayer thickness t for CoFeB(11)/X(t)/Pt(3) systems with $\rm X=Al$ (a), Cr (b), Ta (c) and for CoFeB(11)/Al(t)/Ta(3) (d). The case $\rm t=0$ represents the case of the reference system without interlayer. The red and black lines represent the theoretically expected damping considering reasonable literature values for CoFeB, Al, Cr and Ta. The spin diffusion length of Pt was treated as a variable once for the case without spacer (red) and with spacer (black).}
     \label{damping}
\end{figure}

Fig.~\ref{damping} shows the dependence of $\alpha$ on the interlayer thickness $t$ for CoFeB(11)/X(t)/Pt(3) systems with $\rm X=Al$ (a), Cr (b) and Ta (c). Regardless of the material, all systems show a similar reduction in damping when the interlayer is introduced, which we attribute to the suppression of the MPE. Furthermore, the minimum value of $\alpha$ is similar for all three systems, indicating a minor role of the CoFeB/X interface on damping, except for the CoFeB/Pt interface with MPE.

In Fig.~\ref{damping}\,(a-c) we observe a reduction of $\alpha$ when an interlayer is introduced, which is similar to the observations reported in the literature for other FM/Pt systems~\cite{rojas2014,caminale,mihalceanu}.
In Fig.~\ref{damping}\,(d) the situation is shown for CoFeB/Al(t)/Ta. This system acts as a control series without MPE and provides a reference value for $\alpha_{0}$. In recent reports we have shown that Ta has a very reduced impact on $\alpha$ in CoFeB. 
We observe the lowest values for $\alpha$ in samples with Ta cap. In addition, $\alpha$ does not change when an Al interlayer is introduced. The measured damping for the CoFeB/Ta is only slightly lower than that of the CoFeB/Ta(t)/Pt systems in Fig.~\ref{damping}(c) with a thicker Ta layer. This indicates a reproducible layer quality and implies that for t$_{\rm Ta}=1.6$ nm the spin pumping contribution $\alpha_{\rm sp}$ is negligible.

To further investigate the influence of the magnetic phase in Pt on the damping of the whole sample system, we performed calculations based on a spin relaxation model~\cite{Boone2015}.
Within this model, the initial damping $\alpha_0$ of the bare CoFeB film in the presence of up to two non-magnetic layers is increased by 
\begin{equation} \label{delta_alpha}
\Delta \alpha = \frac{\gamma\hbar^2}{2 e^2 M_s d_\mathrm{FM} } G_\mathrm{eff},
\end{equation}
where $\gamma$ is the gyromagnetic ratio, $\hbar$ the reduced Planck constant, $e$ the electron charge, $M_S$ the saturation magnetization of the ferromagnetic CoFeB layer, and $ d_{FM}$ the thickness of the CoFeB layer. The effective spin mixing conductance including the correction for spin backflow is given by
\begin{equation} \label{G_eff}
2 G_\mathrm{eff} = \frac{2G_{\uparrow\downarrow}}{1+(2G_{\uparrow\downarrow}/G_\mathrm{ext})}.
\end{equation}
$G_{\uparrow\downarrow}$ denotes the bare spin mixing conductance. The spin conductance of the nonmagnetic layer(s) is
\begin{equation} \label{G_ext}
G_\mathrm{ext} = -\frac{\prod_{1,2}}{\prod_{1,1}},
\end{equation}
where $\prod_{i,j}$ are entries from 
\begin{equation} \label{Product_matrix}
\Pi = \prod_{i = 1}^{n} T_i
\end{equation}
with
\begin{equation} \label{T_matrix}
T_i = \begin{bmatrix}
\cosh(\frac{L_i}{\lambda_{\mathrm{spin},i}}) & -\frac{\sigma_i}{\lambda_{\mathrm{spin},i}} \sinh(\frac{L_i}{\lambda_{\mathrm{spin},i}}) \\
  -\frac{\lambda_{\mathrm{spin},i}}{\sigma_i} \sinh(\frac{L_i}{\lambda_{\mathrm{spin},i}}) & \cosh(\frac{L_i}{\lambda_{\mathrm{spin},i}}) \\
  \end{bmatrix}.
\end{equation}\quad

Here, $L_i$ is the thickness of the layer $i$ on top of CoFeB, $\lambda_{\mathrm{spin},i}$ the spin diffusion length and $\sigma_i$ the bulk conductance. We consider a thickness dependent resistivity $\rho_\mathrm{NM}$, with

\begin{equation} \label{roh_thickness}
 \rho_\mathrm{NM} = \rho_{\mathrm{b}} + \frac{\rho_{\mathrm{s}}}{L_\mathrm{NM}}
\end{equation}\quad

where $\rho_{\mathrm{b}}$ is the bulk resistivity (see Table~\ref{valuestable}) of the corresponding element and $\rho_{\mathrm{s}}$ is the interfacial resistivity coefficient. The interfacial resistivity was taken from~\cite{Boone2015} with $3.3\times10^{-16}\,\Omega \rm m^{2}$ for Pt and was also used for the other elements.

The results are shown as lines in Fig~\ref{damping}, representing the theoretically expected evolution of the damping by the spin relaxation considering model. The values of $\sigma_\mathrm{el}$ and $\lambda_\mathrm{spin}$  for CoFeB, Al, Cr and Ta are taken from the literature and are shown in Table~\ref{valuestable}. 
We choose different values for $\lambda_\mathrm{spin}(\mathrm{Pt})$ for CoFeB/Pt and CoFeB/X/Pt to match the measured data. 
Here we distinguish between two cases: The first case, marked in red, shows the $\lambda_\mathrm{spin}(\mathrm{Pt})$ needed to reproduce the total damping of the CoFeB/Pt sample without any spacer layers.
The second case, marked in black, shows the $\lambda_\mathrm{spin}(\mathrm{Pt})$ that would be needed to reproduce the damping evolution of the CoFeB/X/Pt systems. 
As a result, we obtain a $\lambda_\mathrm{spin}(\mathrm{Pt}) \approx 1.0$\,nm in the absence of a spacer layer and a $\lambda_\mathrm{spin}(\mathrm{Pt}) \approx 2.5$\,nm when a spacer layer of Al, Cr or Ta is introduced. Since the higher initial damping of the CoFeB/Pt system can be modeled by a reduced $\lambda_\mathrm{spin}(\mathrm{Pt})$ holding all other parameters constant, we claim that this apparent change in $\lambda_\mathrm{spin}(\mathrm{Pt})$ is caused by the transition of Pt from a phase without a net magnetic moment to a magnetic phase~\cite{Ghosh2012}. The absence or presence of magnetic proximity effects in different samples may contribute to the large range of values for the spin diffusion length of Pt found in the literature. 

\begin{figure}[h]
  \centering
   \includegraphics[width=1\columnwidth]{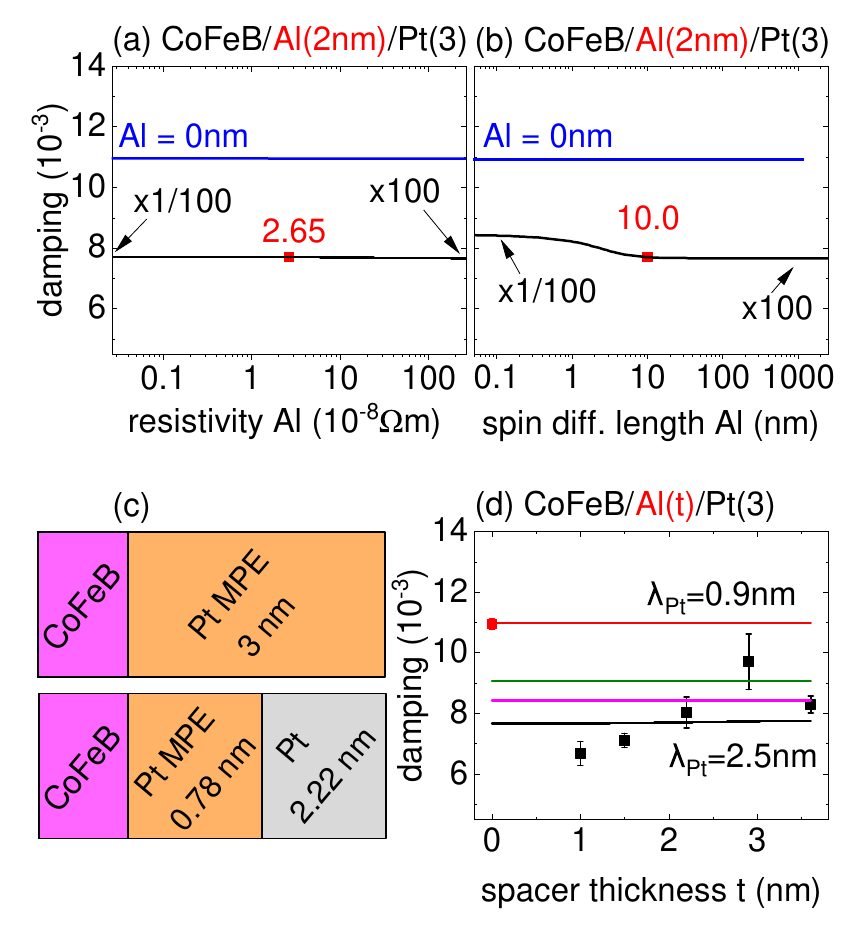}
    \caption{Calculations on the spin relaxation model with variation of resistivity, spin diffusion length and spin mixing conductance. Variation of the resistivity (a) and the spin diffusion length (b) of the Al spacer in CoFeB/Al(2\,nm)/Pt(3\,nm). (c) Sketch of the separation of the homogeneous 3\,nm Pt film magnetized by the MPE in direct contact with CoFeB into a thinner 0.78\,nm magnetized interface layer and a remaining non-magnetized Pt film. (d) Damping data and theoretically expected damping from Fig.~\ref{damping} (red and black line). If the Al spacer is removed and only a 0.78\,nm thick magnetized interfacial layer in Pt is considered, see panel (c), the theoretically expected damping coincides with the black line, within the resolution of the figure. The magenta line is obtained when either reducing the spin diffusion length of Pt to 0.9\,pm or decreasing the Pt resistivity by three orders of magnitude. The green line represents an additional increase in the spin mixing conductance by five orders of magnitude.}
 \label{changing_parameters}
\end{figure}

In addition to the change in the spin diffusion length of Pt, one can consider a change in the resistivity $\sigma$ and the spin diffusion length $\lambda_\mathrm{spin}$ of the spacer layer in contact with the ferromagnet ~\cite{caminale}. In this modeling approach, the extracted $\lambda_\mathrm{spin}(\mathrm{Pt})$ depends on the assumption of a fully magnetized Pt film with a thickness of 3\,nm. This is not consistent with the established MPE mechanism, since the MPE should only magnetize a few monolayers of the neighboring material. 
In the following we want to address this issue as shown in Fig.~\ref{changing_parameters}, using the spin relaxation model.

First, we changed the resistivity and spin diffusion length of the spacer element Al, taking the calculated value of the total damping of CoFeB/Al(2nm)/Pt(3) from Fig.~\ref{damping} as a reference. In Fig.~\ref{changing_parameters}\,(a+b) the expected damping is plotted against the logarithmic change of Al resistivity and spin diffusion length by a factor of 100 to smaller and higher values. It is clearly visible that even such a dramatic change in Al resistivity or spin diffusion length cannot lead to such an enhanced damping as in the absence of the Al spacer layer (blue line). 

Second, we separate the initially fully magnetized homogeneous 3\,nm Pt film (Fig.~\ref{changing_parameters}\,(c) top) into a magnetized two monolayer thick layer of Pt, using the lattice constant of~\cite{HandbookCaP2008} of 3.9\,\si{\angstrom}, and a remaining 2.22\,nm non-magnetized Pt film (Fig.~\ref{changing_parameters}\,(c) bottom). 
Considering these changes, the damping values without Al spacer layer are calculated for varying values of spin diffusion length, resistivity, and spin mixing conductance. These results are shown in Fig.~\ref{changing_parameters}\,(d). The red and black lines again show the calculated damping considering a homogeneously magnetized Pt film, while the case of a magnetized double monolayer Pt film as introduced in Fig.~\ref{changing_parameters}\,(c) bottom leads to an identical curve as the black one within the resolution of the figure. Since the $\lambda_\mathrm{spin}(\mathrm{Pt})$ is higher than the Pt film thickness of 0.78\,nm the introduction of the additional layer does not change the damping in the spin-diffusion model. In the next step we reduced the $\lambda_\mathrm{spin}(\mathrm{Pt})$ to 0.9\,pm (magenta line) to determine the influence of a vanishing spin diffusion length. Compared to the case of a homogeneously magnetized Pt film with $\lambda_\mathrm{spin}(\mathrm{Pt})=0.9 $\,nm (red line), the calculated damping with the interfacially magnetized Pt layer of 0.78\,nm is still too low to explain the measured enhanced damping of the CoFeB/Pt(3) sample. As a last step, we reduced the resistivity $\rho_\mathrm{Pt} $ and increased the spin mixing conductance $ 2G_{\uparrow\downarrow} $ by a factor of $ 10^{3} $ and $ 10^{5} $, respectively. While this reduction of $ \rho_\mathrm{Pt} $ has, in this figure scaling, no effect on the damping (again magenta line), the enhancement of $ 2G_{\uparrow\downarrow} $ leads to a visible increase of the damping (green line), but still both changes are insufficient to describe the enhanced damping of CoFeB/Pt(3).  In summary, we have shown that as soon as we consider only a two-monolayer thin magnetized Pt film with the rest of the 3\,nm film unaffected by MPE, the spin relaxation model is not completely sufficient to explain the observed enhancement of damping in CoFeB/Pt(3), at least by changing only a single parameter.
\\

\begin{table}[]
\caption{\label{valuestable}Bulk conductivities and spin diffusion lengths of Pt, Al, Cr and Ta at room temperature and the spin mixing conductance used for the layer thickness dependent damping calculations from Fig.\ref{damping}.}
\begin{tabular*}{\columnwidth}{@{}p{3pc}p{6.5pc}p{5pc}l@{}}
\toprule
element & $\sigma_\mathrm{el}(10^6 \Omega^{-1}\cdot m^{-1})$ & $\lambda_\mathrm{spin}$(nm) & $G_{\uparrow\downarrow} (10^{15})$\\
\colrule
Pt	&  9.52~\cite{HandbookCaP2008}& MPE-dependent & 5.0\\ 

Al	& 37.7~\cite{HandbookCaP2008}& 10.0~\cite{Bass2007} & 5.0\\ 

Cr & 8.0~\cite{HandbookCaP2008}& 2.1~\cite{Qu2015} & 5.0\\ 

Ta & 7.63~\cite{HandbookCaP2008}& 1.7~\cite{Qu2015} & 0.8\\

\botrule
\end{tabular*}\vspace*{-12pt}
\end{table}

\section{Conclusion}
In conclusion, we have systematically investigated the influence of the MPE in Pt on the spin pumping damping $\alpha$ for CoFeB/Pt bi- and CoFeB/Al,Cr,Ta/Pt trilayer structures.
Independent of the spacer material, the introduction of an interlayer in CoFeB/Pt systems always leads to a reduction of $\alpha$, strongly suggesting that the large damping in CoFeB/Pt originates in the magnetically polarized Pt layer caused by the MPE. The presence of MPE in Pt was experimentally verified by table-top element sensitive HHG-TMOKE spectroscopy.

Furthermore, the phenomenological spin transport model reproduced the changes of $\alpha$ in the presence of MPE in Pt by a reduction of the spin diffusion length of the whole Pt film. However, the application of the spin transport model to a partially magnetically polarized Pt film shows the limitations of the model to describe the observed damping behavior by changing one or more spacer and Pt parameters.

\begin{acknowledgements}
    This work was funded by the Deutsche Forschungsgemeinschaft (DFG, German Research Foundation) - TRR 173 - 268565370 Spin + X: spin in its collective environment (Project A08, B04 and B13), the M-era.Net through the HEUMEM project and the Carl Zeiss Stiftung. B.S. further acknowledges funding by the Dynamics and Topology Research Center (TopDyn) funded by the State of Rhineland Palatinate.
\end{acknowledgements}

\end{document}